\documentclass[english, aps, preprint, superscriptaddress]{revtex4}
\usepackage{lmodern}
\usepackage[T1]{fontenc}
\usepackage[latin9]{inputenc}
\usepackage{color}
\usepackage{verbatim}
\usepackage{graphicx}
\usepackage{amssymb}
\usepackage{esint}



\usepackage{babel}

\begin{document}

\title{Scaling laws of resistive magnetohydrodynamic reconnection in the
high-Lundquist-number, plasmoid-unstable regime}

\author{Yi-Min Huang }

\email{yimin.huang@unh.edu}

\affiliation{Space Science Center, University of New Hampshire, Durham, NH 03824}

\affiliation{Center for Integrated Computation and Analysis of Reconnection and
Turbulence}

\affiliation{Center for Magnetic Self-Organization in Laboratory and Astrophysical
Plasmas}

\author{A. Bhattacharjee }

\email{amitava.bhattacharjee@unh.edu}

\affiliation{Space Science Center, University of New Hampshire, Durham, NH 03824}

\affiliation{Center for Integrated Computation and Analysis of Reconnection and
Turbulence}

\affiliation{Center for Magnetic Self-Organization in Laboratory and Astrophysical
Plasmas}
\begin{abstract}
The Sweet-Parker layer in a system that exceeds a critical value of
the Lundquist number ($S$) is unstable to the plasmoid instability.
In this paper, a numerical scaling study has been done with an island
coalescing system driven by a low level of random noise. In the early
stage, a primary Sweet-Parker layer forms between the two coalescing
islands. The primary Sweet-Parker layer breaks into multiple plasmoids
and even thinner current sheets through multiple levels of cascading
if the Lundquist number is greater than a critical value $S_{c}\simeq4\times10^{4}$.
As a result of the plasmoid instability, the system realizes a fast
nonlinear reconnection rate that is nearly independent of $S$, and
is only weakly dependent on the level of noise. The number of plasmoids
in the linear regime is found to scales as $S^{3/8}$, as predicted
by an earlier asymptotic analysis (Loureiro \emph{et al.}, Phys. Plasmas
\textbf{14}, 100703 (2007)). In the nonlinear regime, the number of
plasmoids follows a steeper scaling, and is proportional to $S$.
The thickness and length of current sheets are found to scale as $S^{-1}$,
and the local current densities of current sheets scale as $S^{-1}$.
Heuristic arguments are given in support of theses scaling relations. 
\end{abstract}
\maketitle

\section{Introduction}

Recent studies of nonlinear reconnection in large high-Lundquist-number
($S$) plasmas, based on resistive magnetohydrodynamics (MHD) \cite{BhattacharjeeHYR2009}
as well as fully kinetic simulations that include a collision operator
\cite{DaughtonRAKYB2009} have produced a surprise. It is seen in
these studies that above a critical value of the Lundquist number,
the system deviates qualitatively from the predictions of Sweet-Parker
theory \cite{Sweet1958,Parker1963} which has been the standard model
for reconnection in the high-$S$ regime. In the Sweet-Parker model,
the reconnection layer has the structure of Y-points, with a length
of the order of the system size, and a width given by $\delta_{SP}=L/S^{1/2}$,
where $S=LV_{A}/\eta$ is the Lundquist number based on the system
size $L$, the Alfv\'en speed $V_{A}$, and the magnetic diffusivity
$\eta$. The Sweet-Parker model is usually considered to be a model
of \textquotedblleft{}slow\textquotedblright{} reconnection as it
predicts the reconnection rate to scale as $S^{-1/2}$. (The Petschek
model \cite{Petschek1964} predicts a much weaker dependence on $S$,
with the maximum reconnection rate $\sim1/\log S$. However, it has
become clear over the years that Petschek model is realizable only
when the resistivity is locally enhanced around the reconnection site.
\cite{Biskamp2000}) For weakly collisional systems such as the solar
corona, the Lundquist number is typically very large ($\sim10^{12}-10^{14}$)
and the Sweet-Parker reconnection time scale is of the order of years,
much too slow to account for fast events such as solar flares. The
Sweet-Parker model is based on the assumption of the existence of
a long thin current layer. Although it has been known for some time
that such a thin current layer may be unstable to a secondary tearing
instability (referred to hereafter as the plasmoid instability) which
generates plasmoids,\cite{BulanovSS1979,LeeF1986,Biskamp1986,YanLP1992,ShibataT2001,Lapenta2008}
it has been realized only fairly recently that the Sweet-Parker layer
actually becomes more unstable as the Lundquist number increases,
with a linear growth rate $\gamma\sim S^{1/4}$ and the number of
plasmoids $\sim S^{3/8}$.\cite{LoureiroSC2007,BhattacharjeeHYR2009}
In a recent paper (Ref. \cite{BhattacharjeeHYR2009}, hereafter referred
to as Paper I), Bhattacharjee\emph{ et al}. have presented numerical
results that suggest strongly that as a consequence of the plasmoid
instability, the system evolves into a nonlinear regime in which the
reconnection rate becomes weakly dependent on $S$.

The primary goal of this paper is to strengthen the results obtained
in Paper I in two significant ways: first, to present new simulation
results with a modified initial condition that enables us to obtain
stronger scaling results on the nonlinear reconnection rate and the
number of plasmoids generated in the nonlinear regime, and second,
a simple heuristic model that is consistent with the results of the
simulation and fortifies the claim in Paper I that the reconnection
rate in the nonlinear regime of the plasmoid instability is fast and
independent of $S$.

\section{Numerical Model}

The initial condition in Paper I does not have a thin current sheet
to begin with. It has four magnetic islands and is unstable to an
ideal coalescence instability. After the onset of the coalescence
instability, a Sweet-Parker current sheet is created when two islands
are attracted toward each other.\cite{LongcopeS1993} In this case,
there is a relatively long initial transient period before the reconnection
process starts. Furthermore, the dynamics of the system are complicated
by the sloshing of coalescing islands that causes the primary Sweet-Parker
layer to lengthen or shorten from time to time. This makes it difficult
to verify the predictions of linear theory in this particular system.
The present study seeks remedies to these two drawbacks. We still
consider the merging of two islands, but now put them in close contact
initially. There is an initial current layer between the flux tubes,
which quickly (typically within less than one Alfv\'en time) adjusts
its width depending on the Lundquist number to form the primary Sweet-Parker
layer, which may subsequently become unstable to the plasmoid instability.
In this new system, the transient period is significantly shortened
and the sloshing between islands is largely eliminated. It is still
not easy to measure the linear growth rate in this new model, but
we can at least verify the scaling of the number of plasmoids in the
linear regime. 

\begin{figure}
\begin{centering}
\includegraphics[scale=0.8]{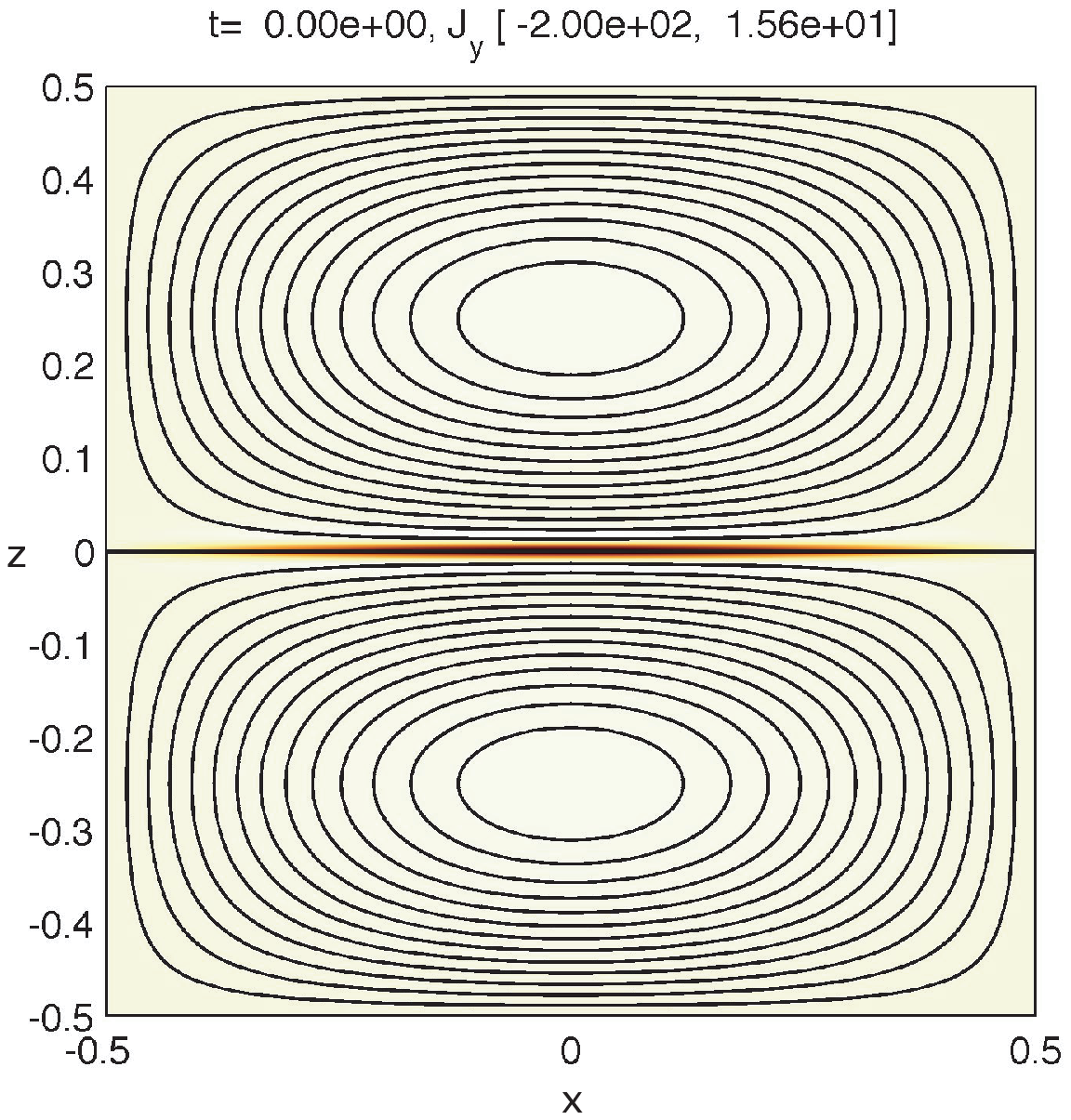}
\par\end{centering}

\caption{(Color online) The initial current density, with magnetic field lines
(constant $\psi$ contours) overlaid. \label{fig:initial_condition}}

\end{figure}

The initial condition is similar to the model of Uzdensky and Kulsrud.\cite{UzdenskyK2000}
To start with, let $\psi_{0}=\cos\left(\pi x\right)\sin\left(2\pi\left|z\right|\right)/2\pi$
and $\mathbf{B}_{0}=\nabla\psi_{0}\times\mathbf{\hat{y}}$ in the
domain $(x,z)\in[-1/2,1/2]\times[-1/2,1/2]$. The $\psi_{0}$ so defined
satisfies $\nabla^{2}\psi_{0}=-5\pi^{2}\psi_{0}$. If the pressure
is set to $p_{0}=C+5\pi^{2}\psi_{0}^{2}/2,$ with $C$ an arbitrary
constant, the system is in force balance. However, the magnetic field
defined by $\psi_{0}$ has a tangential discontinuity at $z=0$, which
causes numerical difficulties. We smooth it out as $\psi=\tanh\left(\alpha z\right)\cos\left(\pi x\right)\sin\left(2\pi z\right)/2\pi,$
where $\alpha$ is a large number. This smoothed function $\psi$
no longer satisfies $\nabla^{2}\psi=f(\psi)$ and the magnetic force
cannot be balanced entirely by pressure. However, the magnetic force
can be canceled to a large extent if the pressure is set to $p=p_{0}+(B_{x0}^{2}-B_{x}^{2})/2$,
where $B_{x0}=-\partial_{z}\psi_{0}$ and $B_{x}=-\partial_{z}\psi$,
respectively. We assume the isothermal equation of state, $p=2\rho T$,
in our simulation. We choose $T=3$, $C=2T$, and $\alpha=100$. For
these parameters, the initial plasma density varies from $0.96$ to
$1.1$ and the plasma beta ($\beta\equiv p/B^{2}$) obeys the inequality
$\beta\gtrsim6$. The system is, therefore, nearly incompressible.
Figure \ref{fig:initial_condition} shows the initial current density
and magnetic field lines.

We find in the present study that the plasmoid instability depends
on the noise level of the system, at least when the Lundquist number
is not far above the critical value. Due to the outflow in the primary
Sweet-Parker layer, if the noise level is low, the plasmoid instability
may not grow to visible size before being convected out. When we seed
the system initially with random noise, at relatively
low values of the Lundquist number ($\sim10^{5}$) we obtain a short
burst of plasmoids, following which the current layer becomes stable
again after all the plasmoids are convected out. Because of this,
we have included a random forcing in the system, which enables us
also to study the effect of noise level on the reconnection rate.
(Notice that we did not apply any random forcing or noise in Paper
I. The sloshing between coalescing islands is a natural source of
noise, but uncontrolled.) The governing equations for the time evolution
of the system are: \begin{equation}
\partial_{t}\rho+\nabla\cdot\left(\rho\mathbf{v}\right)=0,\label{eq:2}\end{equation}
\begin{equation}
\partial_{t}(\rho\mathbf{v})+\nabla\cdot\left(\rho\mathbf{vv}\right)=-\nabla p-\nabla\psi\nabla^{2}\psi+\epsilon\mathbf{f}(\mathbf{x},t),\label{eq:3}\end{equation}
\begin{equation}
\partial_{t}\psi+\mathbf{v}\cdot\nabla\psi=\eta\nabla^{2}\psi,\label{eq:4}\end{equation}
where a random forcing term $\epsilon\mathbf{f}(\mathbf{x},t)$ is
added to the right hand side of the momentum equation (\ref{eq:3}).
The forcing function is white noise in both space and time with $\left\langle \mathbf{f}\right\rangle =0$,
and $\epsilon$ is a small parameter which controls the noise level.
By white noise we mean that $\left\langle f_{i}(\mathbf{x},t)f_{j}(\mathbf{x'},t')\right\rangle \sim\delta_{ij}\delta(\mathbf{x}-\mathbf{x}')\delta(t-t'),$
where $i$ and $j$ can be $x$ or $z$ and $\left\langle \,\right\rangle $
is the ensemble average. Care has to be taken to ensure that the discrete
representation is independent of the time step. Our implementation
is similar to that in Ref.\cite{Alvelius1999}. It is convenient
to set $\mathbf{f}=\rho\mathbf{a}$, where $\mathbf{a}$ is a random
acceleration. In a single time step, the momentum density evolves
from $\rho\mathbf{v}$ to $\rho\mathbf{v}+\epsilon\rho\mathbf{a}\Delta t$
(neglecting other forces), and the kinetic energy density evolves
from $\rho v^{2}/2$ to $\rho v^{2}/2+\epsilon\rho\mathbf{a}\cdot\mathbf{v}\Delta t+\epsilon^{2}\rho a^{2}\Delta t^{2}/2$.
Therefore, the average power density from the random force is \[
\frac{\left\langle \epsilon\rho\mathbf{a}\cdot\mathbf{v}\Delta t+\epsilon^{2}\rho a^{2}\Delta t^{2}/2\right\rangle }{\Delta t}=\frac{1}{2}\epsilon^{2}\rho\Delta t\left\langle a^{2}\right\rangle ,\]
where $\left\langle \mathbf{a}\cdot\mathbf{v}\right\rangle =0$ is
used, and $a^{2}=\mathbf{a\cdot a}$. At each grid point, we set $a_{x}$,
$a_{z}$ to random numbers between $-1$ and $1$ with a uniform probability
distribution, divided by $\sqrt{\Delta t}$. That is,\[
a_{i}=\frac{\mbox{rand}(-1,1)}{\sqrt{\Delta t}}.\]
Then $\Delta t\left\langle a^{2}\right\rangle =2/3$ and is independent
of $\Delta t$. The average power density is $\epsilon^{2}\rho/3$
and the total power $\epsilon^{2}M/3,$ with $M$ the total mass ($M\simeq1$
in our simulation). We use $\epsilon=10^{-5}-10^{-3}$ in our simulations
and the corresponding power density ranges from $3\times10^{-11}$
to $3\times10^{-7}$. Since our simulations typically last only a
few Alfv\'en times, the energy input from random forcing is negligible
compared to the total magnetic energy ($\sim O(1)$) in the system.
This ensures that the random forcing only provides noise for the instability
to grow but does not otherwise alter the system in a significant way
(see more discussion in Sec. \ref{sec:Numerical-Results}).

Our numerical algorithm \cite{GuzdarDMHL1993} uses finite differences
with a five-point stencil in each direction, and a second-order accurate
trapezoidal leapfrog method for time stepping. We use a uniform grid
along the $x$ direction, and a nonuniform grid in the $z$ direction
that packs high resolution around $z=0$ in order to resolve the sharp
spatial gradients in the reconnection layer. Perfectly conducting
($\partial_{t}\psi=0$), impenetrable ($\mathbf{v}\cdot\mathbf{\hat{n}}=0$),
and free slipping ($\mathbf{\hat{n}}\cdot\nabla$$\left(\mathbf{\hat{n}}\times\mathbf{v}\right)=0$)
boundary conditions are assumed ($\mathbf{\hat{n}}$ is the unit normal
vector to the boundary). We further assume reflection symmetry along
the $x$ axis and only the region $z\ge0$ is simulated. The
highest resolution is $12800$ in $x$ and $1600$ in $z$, with the
smallest grid size $\Delta z=5.4\times10^{-6}$.

\section{Numerical Results\label{sec:Numerical-Results}}

One of the key objectives of this study is to determine the scaling
of reconnection rate in the plasmoid-unstable regime. To quantify
the speed of reconnection, we measure the time it takes to reconnect
a significant portion of the magnetic flux within the two merging
islands. The amount of magnetic flux in an island is $\psi_{max}-\psi_{s}$,
where $\psi_{max}$ is the maximum of $\psi$ in the island and $\psi_{s}$
is the value of $\psi$ at the separatrix separating the two merging
islands. Initially $\psi_{max}\simeq0.16$ and it remains approximately
unchanged since the resistivity is low; therefore it suffices to just
measure $\psi_{s}$. We denote the time it takes to reconnect from
$\psi_{s}=0.01$ to $\psi_{s}=0.05$ as $t_{rec}$. The starting point
$\psi_{s}=0.01$ is chosen to allow the plasmoid instability to build
up, and the end point $\psi_{s}=0.05$ is chosen such that the reconnection
layer does not shorten too much compared with that in the initial
condition. The range corresponds to reconnecting $25\%$ of the initial
flux. 

\begin{figure}
\begin{centering}
\includegraphics[scale=0.8]{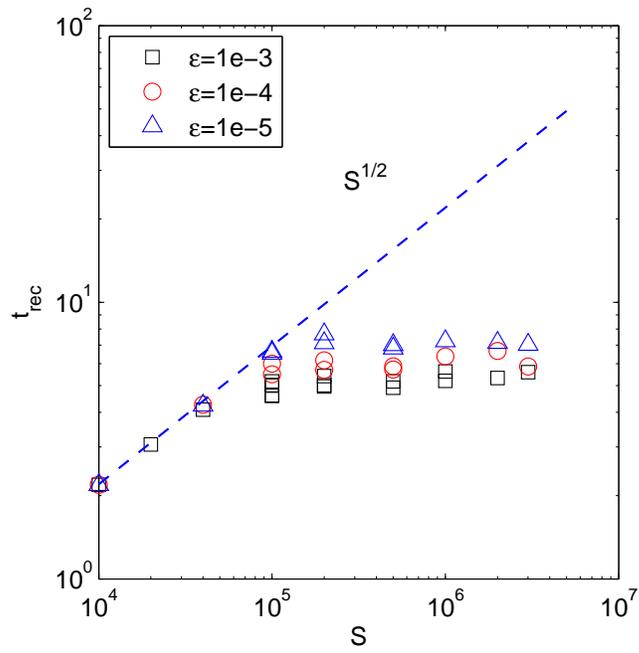}
\par\end{centering}

\caption{(Color online) The reconnection time $t_{rec}$ for various $S$ and
$\epsilon$. The dashed line is the Sweet-Parker scaling.\label{fig:reconnection-time}}

\end{figure}

Figure \ref{fig:reconnection-time} shows the reconnection time $t_{rec}$
for various $S$ and $\epsilon$. For lower $S$, the reconnection
time $t_{rec}\sim S^{1/2}$, as expected from the Sweet-Parker theory.\cite{UzdenskyK2000}
The critical Lundquist number for plasmoid instability is about $S_{c}\simeq4\times10^{4}$.
Above $S_{c}$, the reconnection time $t_{rec}$ deviates from the
Sweet-Parker scaling and becomes shorter. In the plasmoid unstable regime, the reconnection
time is nearly independent of $S$. However, the reconnection time
has a weak dependence on the noise level throughout the $S$ range
we have tested. The plateaued values of $t_{rec}$ in the high-$S$
regimes for $\epsilon=10^{-3}$, $10^{-4}$, and $10^{-5}$ are $5.30\pm0.27$,
$6.10\pm0.41$, and $7.05\pm0.16$, respectively. Here we take the
average values over the range $S=5\times10^{5}$ to $3\times10^{6}$,
and the errors represent the standard deviation. We have tested the
convergence of our numerical results by varying the resolution, the
time step, and the random seed for selected runs. These are represented
by multiple data points with the same parameters in Figure \ref{fig:reconnection-time}.
The results are fairly consistent, with fluctuations no more than
a few percent. The dependence of $t_{rec}$ on $\epsilon$ may be
tentatively fit with a power law, which gives $t_{rec}\sim\epsilon^{-0.062}$.
However, given the limited range of the parameter space we have explored,
this scaling should not be considered as conclusive. 

The global characteristic values for $V_{A}$ and $B$ are about $1$,
which yield the normalized average reconnection rate as \[
\frac{1}{BV_{A}}\left\langle \frac{d\psi_{s}}{dt}\right\rangle =\frac{0.04}{t_{rec}}.\]
In the high Lundquist number regime, $t_{rec}\simeq5$
to $7$ from our simulations and the normalized reconnection rate
is in the range $0.006$ to $0.008$. The normalized reconnection
rate obtained here is similar to the result of Paper I. 

\begin{figure}
\begin{centering}
\includegraphics[scale=0.8]{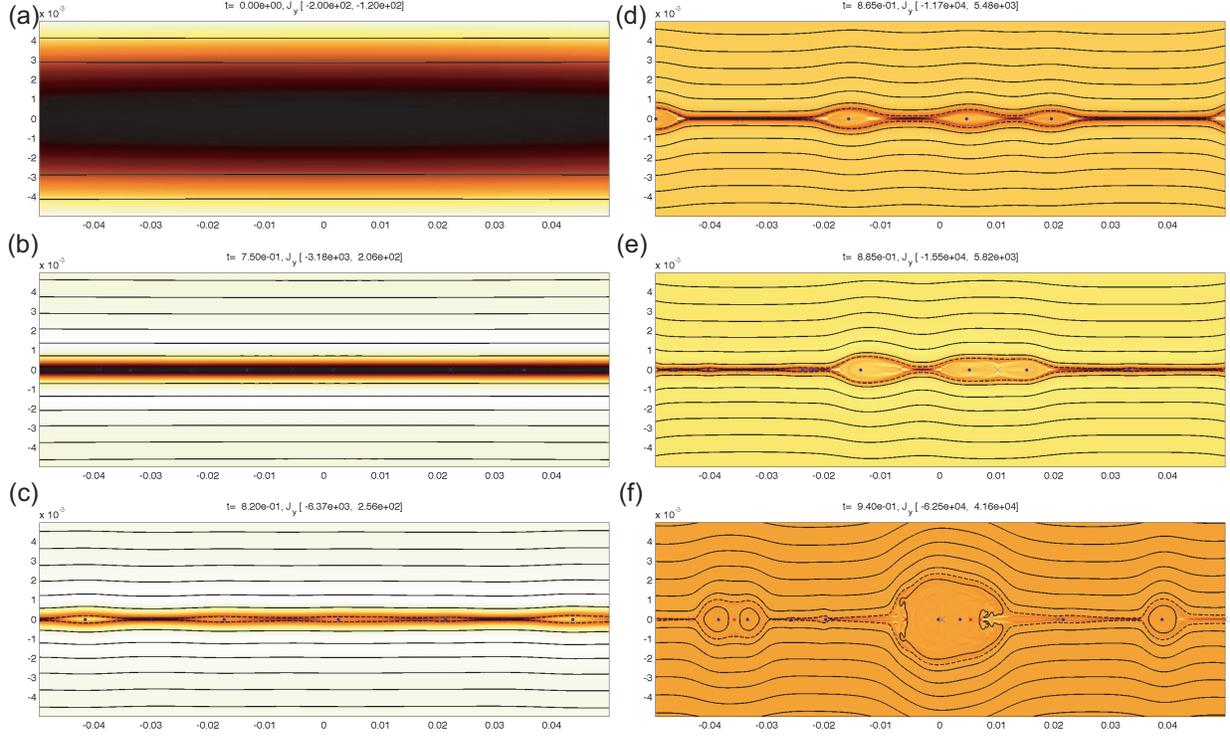}
\par\end{centering}

\caption{(Color online) Time sequence of the current density, overlaid with
magnetic field lines, for the case $S=3\times10^{6}$, $\epsilon=10^{-3}$.
The dashed line indicates the separatrix. The dots and crosses indicate
O points and X points, respectively. The blue and the red colors indicate
the two types of X and O points (blue for type I and red for type
II) . Notice that the vertical axis is stretched for better visualization.
\label{fig:timeseq}}

\end{figure}

\begin{figure}
\begin{centering}
\includegraphics[scale=0.8]{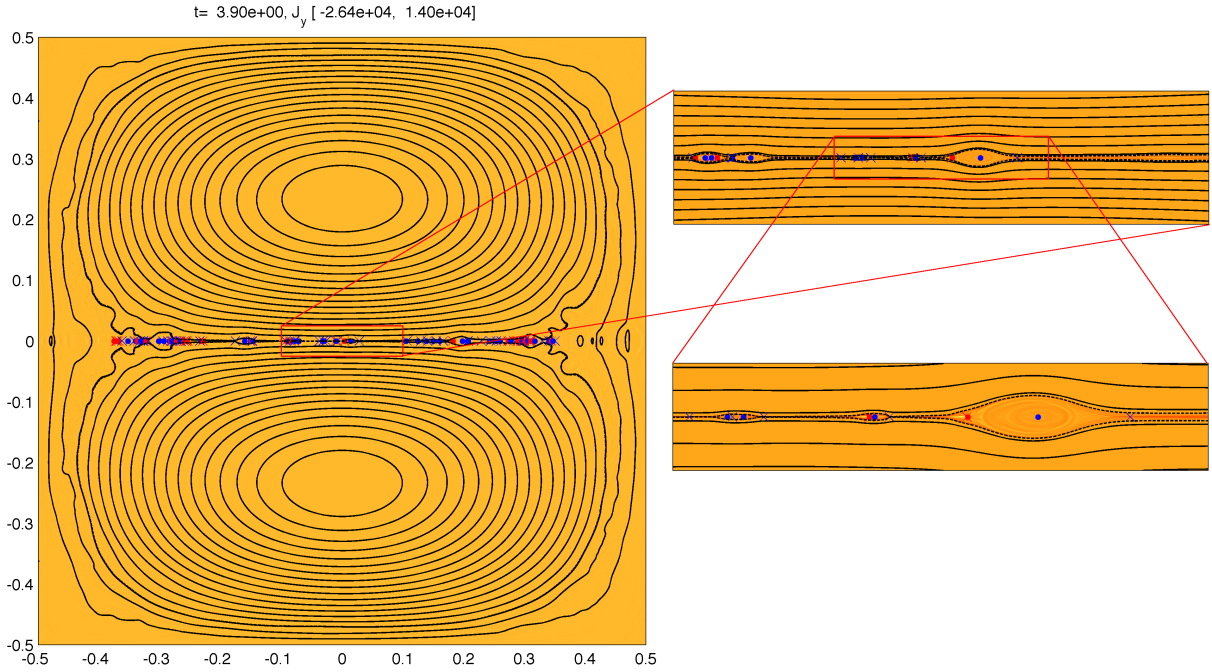}
\par\end{centering}

\caption{(Color online) The global configuration at a later time, $t=3.9$.
Show on the right are close-ups of the reconnection layer.\label{fig:global} }

\end{figure}

Figure \ref{fig:timeseq} shows a time sequence of the current density,
overlaid with magnetic field lines, within a small area ($1/1000$)
of the whole domain for a case with $\epsilon=10^{-3}$ and $S=3\times10^{6}$.
The initial current layer (panel (a)) quickly thins down to form the
primary Sweet-Parker layer (panel(b)), which becomes unstable to the
plasmoid instability (panel (c)). As the instability proceeds, the
plasmoids grow in size and the current sheets between plasmoids are
again Sweet-Parker like (panel (d)). These secondary Sweet-Parker
current sheets are thinner than the primary one and are again unstable
to the tertiary plasmoid instability (panel (e)). This process of
multiple stages of cascading resembles the scenario envisaged in Ref.\cite{ShibataT2001}.
The plasmoids can merge to form larger ones and new plasmoids are
constantly generated (panel (f)). Figure \ref{fig:global}
shows the global configuration at a later time $t=3.9$, as well as
close-ups of the reconnection layer. The figure shows that on a large,
coarse-grained scale, the configuration looks Sweet-Parker like, except
for the important difference that the reconnection layer is no longer
a single extended current sheet, but is made up of a sequence of copious
plasmoids and current sheets. 

\begin{figure}
\begin{centering}
\includegraphics[scale=0.9]{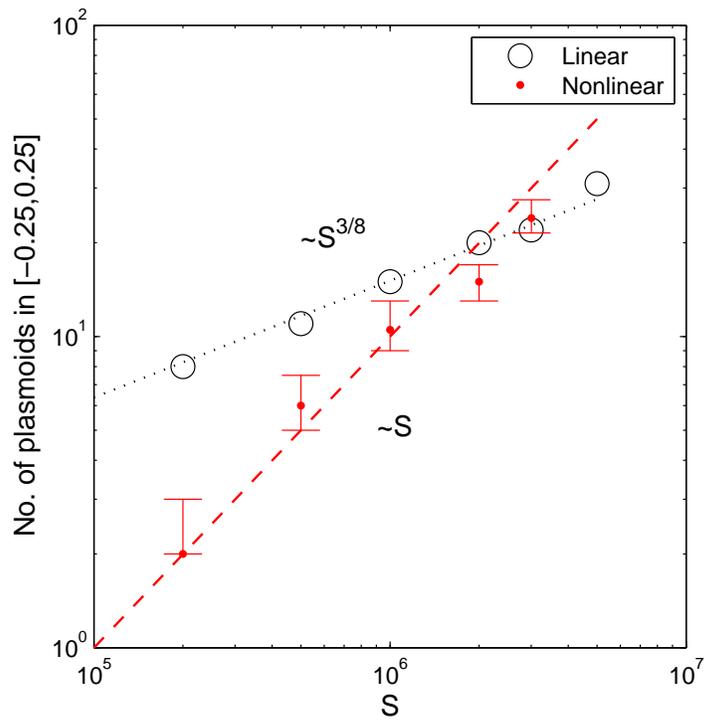}
\par\end{centering}

\caption{(Color online) Number of plasmoids in the linear and the nonlinear
regime. \label{fig:Number-of-plasmoids}}

\end{figure}

Linear theory predicts that the number of plasmoids, $n_{p}^{L}$,
scales as $S^{3/8}$.\cite{LoureiroSC2007,BhattacharjeeHYR2009}
We verify this by counting the maximum number of plasmoids within
the central region, $-0.25\le x\le0.25$, before the plasmoid instability
becomes highly nonlinear (roughly corresponds to panel (c) in Figure
\ref{fig:timeseq}). Figure \ref{fig:Number-of-plasmoids} shows the
number of plasmoids versus $S$, for $\epsilon=10^{-3}$, in both
linear and nonlinear (see the discussion later) regimes. The result
in the linear regime is in good agreement with the $S^{3/8}$ scaling
predicted by asymptotic analysis. This scaling has been verified by
Samtaney \emph{et al}. with local simulations up to $S=10^{8}$.\cite{SamtaneyLUSC2009} 

In the fully nonlinear regime, the plasmoid dynamics are very complicated
and constantly evolving. Plasmoids may grow in size, coalesce with
each other to form larger plasmoids, and finally get ejected into
the downstream region. Meanwhile, new plasmoids are constantly generated
in the reconnection layer. We may regard the reconnection layer with
multiple plasmoids as a statistical steady state. As a simple, first
approximation, we expect the cascading to stop when the current sheet
segments between plasmoids become stable. We may imagine the reconnection
layer as a chain of plasmoids connected by marginally stable Sweet-Parker
current sheets. For given $\eta$ and $V_{A}$,
the critical length of a marginally stable current layer is $L_{c}=S_{c}\eta/V_{A}$.
Therefore we expect the number of plasmoids in the nonlinear regime,
$n_{p}^{NL}$, to scale like $L/L_{c}\sim S/S_{c}$. Furthermore,
the thickness of each Sweet-Parker sheet is $\delta_{c}\sim L_{c}/\sqrt{S_{c}}\sim\eta\sqrt{S_{c}}/V_{A}\sim\delta_{SP}\sqrt{S_{c}/S}$,
and the current density $J\sim B/\delta_{c}\sim BV_{A}/\eta\sqrt{S_{c}}\sim BS/L\sqrt{S_{c}}$
. If we identify the reconnection rate with $\eta J$, then the reconnection
rate $\sim\eta J\sim BV_{A}/\sqrt{S_{c}}$, which is independent of
$S$. This is consistent with our finding that the reconnection rate
is nearly independent of $S$ in the high-$S$ regime. Clearly, the
assumption that all current sheets are marginally stable, and therefore
all identical, is oversimplified. If we look at
the individual current sheets, there are a whole variety of them,
each with a different length, width, and current density. Therefore,
the system is better described with a statistical approach. If we
neglect complications such as asymmetry or background shear flow,
and consider the simple Sweet-Parker picture for each current sheet,
then the local Lundquist number $S_{local}\equiv V_{A}l/\eta$ is
the only dimensionless parameter associated with it. Here $l$ denotes
the length of the current sheet. The current sheet thickness will
be $\delta\sim l/\sqrt{S_{local}}$, as predicted by the Sweet-Parker
theory. The local Lundquist number being greater or smaller than $S_{c}$
determines whether a current sheet may or may not further break into
plasmoids and even smaller current sheets. Because it is the local
Lundquist number that determines the cascading of a local current
sheet to even smaller scales, we hypothesize that the probability
distribution of $S_{local}$ is independent of the global Lundquist
number. The underlying assumption is that, if we consider the ensemble
of local current sheets and characterize each current sheet by a dimensionless
parameter $S_{local}$, there is a similarity across systems of different
global Lundquist numbers. If we further assume that the local upstream
Alfv\'en speed is determined by global conditions, it follows that
statistically the length $l$ and the thickness $\delta$ of a current
sheet scale as $\eta$ , and the current density $J$ scales as $\eta^{-1}$.
If we consider the simple picture that two neighboring current sheets
are separated by a plasmoid, then $l\sim\eta$ implies the number
of plasmoids in the nonlinear regime $\sim\eta^{-1}$. 

Now we proceed to examine whether the conclusions
from the simple heuristic argument are consistent with our simulation
data. Here we present the results from cases with $\epsilon=10^{-3}$.
Results from other values of $\epsilon$ are similar. We count the
number of plasmoids by first identifying X-points and O-points along
$z=0$. There are two types of O-points, the local minimum (type I)
and the local maximum (type II) of $\psi$. Likewise, there are two
types of X-points, one with $\partial_{x}^{2}\psi<0$, $\partial_{z}^{2}\psi>0$
(type I) and the other with $\partial_{x}^{2}\psi>0$, $\partial_{z}^{2}\psi<0$
(type II). In the linear regime, only X-points and O-points of type
I are present. When plasmoids start to coalesce with each other, type
II null points may be created. We count the number of type I O-points
within $-0.25\le x\le0.25$ as the number of plasmoids in the nonlinear
regime. As shown in Figure \ref{fig:Number-of-plasmoids}, the number
of plasmoids in the nonlinear regime appears to agree with the $\sim S$
scaling. Because the number of plasmoids fluctuates, the median value
is used; the error bar indicates the range between the first and the
third quartiles. Notice that although the number of plasmoids in the
nonlinear regime follows a steeper scaling than that in the linear
regime, it is not until about $S=2\times10^{6}$ that the nonlinear
scaling catches up with the linear counterpart. This is because at
lower $S$, the coalescence and ejection of plasmoids exceeds the
generation of new plasmoids. Equating the estimate, $n_{p}^{L}\simeq S^{3/8}/2\pi$,
from the linear theory \cite{LoureiroSC2007,BhattacharjeeHYR2009}
with the heuristic nonlinear estimate, $n_{p}^{NL}\simeq S/S_{c}$,
and using $S_{c}\simeq4\times10^{4}$, we obtain $n_{p}^{L}\simeq n_{p}^{NL}$
when $S\simeq1.2\times10^{6}$, which is in approximate agreement
with the observed $S=2\times10^{6}$. 

To examine the statistics of current sheets, we
have to first set up a diagnostic for a current sheet, which is subject
to a certain degree of arbitrariness. We search for local maxima of
$J$ greater than $10\%$ of the global maximum $J_{max}$ within
$-0.25\le x\le0.25$ as potential sites of current sheets. However,
two neighboring maxima are regarded as separate current sheets only
when the trough between them is lower than $25\%$ of the greater
of the two. The length $l$ and the thickness $\delta$
of a current sheet are measured by the locations where the current
density drops to $25\%$ of the local maximum $J$ of the current
sheet. 

\begin{figure}
\begin{centering}
\includegraphics[scale=0.8]{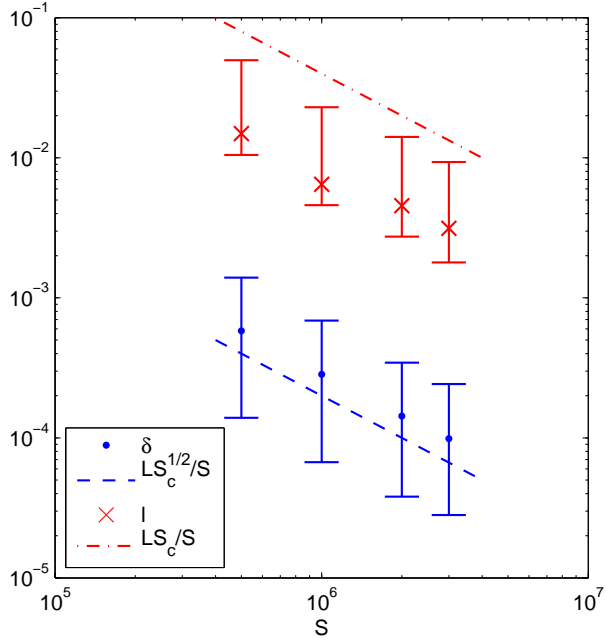}
\par\end{centering}

\caption{(Color online) Scalings of the current sheet thickness $\delta$ and
length $l$ with respect to the global Lundquist number $S$. Also
shown for reference are the predictions from the heuristic argument.\label{fig:Scalings-Length}}

\end{figure}

\begin{figure}
\begin{centering}
\includegraphics[scale=0.8]{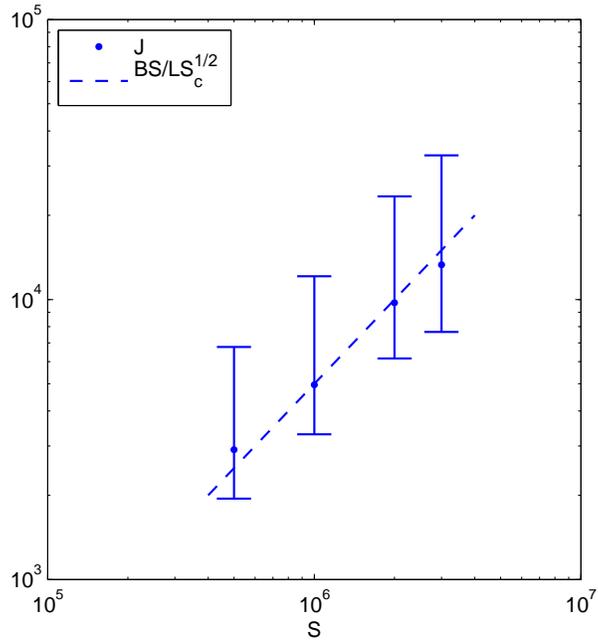}
\par\end{centering}

\caption{(Color online) The scaling of the local peak current density. Also
shown for reference is the prediction from the heuristic argument.
\label{fig:Scaling-J} }

\end{figure}

Figures \ref{fig:Scalings-Length} and \ref{fig:Scaling-J}
show scalings of the thickness $\delta$, length $l$ , and current
density $J$ with respect to the global Lundquist number $S$. The
data are collected from time slices during the period to reconnect
$25\%$ of the flux in each case. Again the median values are used,
and error bars indicate the range between the first and the third
quartiles. Also shown for reference are the predictions from the heuristic
argument based on marginally stable current sheets, i.e., $l\sim L_{c}\sim LS_{c}/S$,
$\delta\sim\delta_{SP}(S_{c}/S)^{1/2}\sim LS_{c}^{1/2}/S$, and $J\sim B/\delta\sim BS/LS_{c}^{1/2}$.
It is evident that the characteristics of current sheets are distributed
over a broad range, as indicated by the rather large error bars. Clearly,
the observed quantities follow the expected scalings. Quite surprisingly,
the predictions for $\delta$ and $J$ from the heuristic argument
are in good agreement with the observed median values, even though
the argument itself is rather crude. However, the prediction of $l$
appears to be systematically an overestimate, and lies at the larger
end of the numerically observed lengths. This is consistent with the
fact that $L_{c}=LS_{c}/S$ is the critical length just above which
the plasmoid instability is triggered. One may wonder how the prediction
of $\delta$ can be consistent with the observed values when the prediction of $l$ is an overestimate. A possible explanation
is that, the heuristic argument assumes Sweet-Parker-like local current
sheets, but clearly not all current sheets in simulations are Sweet-Parker-like.
The existence of non-Sweet-Parker-like current sheets is evident from
the movie available online, which is for the case $S=3\times10^{6}$,
$\epsilon=10^{-3}$. 

\begin{figure}
\begin{centering}
\includegraphics[scale=0.6]{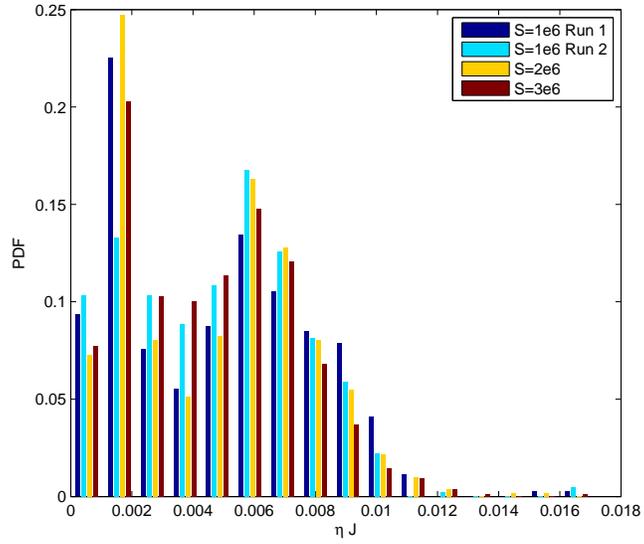}
\par\end{centering}

\caption{(Color online) The probability distribution function of local $\eta J$
for different $S$, from cases with $\epsilon=10^{-3}$. \label{fig:Distribution}}

\end{figure}

Let us now take a more detailed look into the statistics
of current sheets. Figure \ref{fig:Distribution} shows the probability
distribution functions (PDFs) of $\eta J$ for $S=10^{6},\,2\times10^{6},\,3\times10^{6}$,
from cases with $\epsilon=10^{-3}$. The case $S=10^{6}$ has been
done with two runs. The PDFs of $\eta J$ from different runs clearly
show a degree of similarity, which lends some support to our hypothesis
of similarity across systems of different global Lundquist number.
However, we also notice some differences between the PDFs from different
runs. Even the two runs with $S=10^{6}$ show a significant variation
in the PDFs. Therefore, more study is needed to further assess the
validity of our hypothesis. Ideally the same global setting should
be repeated many times with different random seeds for better statistics,
but that is computationally too expensive to be done at the present
time.

Before we conclude this Section, we remark on a few
subtle issues. In the heuristic argument given above, we have used
the quantity $\eta J$ to estimate the reconnection rate. Strictly
speaking, this is valid when the X-point and the stagnation point
of the flow coincide, which is not necessarily the case when the reconnection
layer is embedded with multiple X-points and plasmoids. Notwithstanding
this caveat, we generally find that the peak value of $\eta J$ is
a reasonable measure of the reconnection rate.

We also address the issue of whether random forcing,
by itself, can significantly enhance the reconnection rate. An estimate
of the effect of reconnection rate due to random fluctuations is the
quantity $\left|\tilde{\mathbf{v}}\times\mathbf{B}\right|$ at the
reconnection layer, where $\tilde{\mathbf{v}}$ is the random velocity
fluctuation. Our estimates indicate that the contribution of random
fluctuations is less than 1\% of the observed reconnection rate in
the high-$S$ regime. This conclusion is reinforced by the fact that
in the plasmoid stable regime, the variations in $t_{rec}$ for different
$\epsilon$ are negligible. This is qualitatively different from a
recent turbulent magnetic reconnection study by Loureiro \emph{et
al.},\cite{LoureiroUSCY2009} where the system
is more strongly driven, and the reconnection rate shows a noticeable
dependence on the magnitude of the forcing even for a Lundquist number
as low as $10^{3}$.

\section{Summary and Conclusion}

In summary, we have shown through a series of simulations that resistive
MHD can achieve a fast reconnection rate in the high-Lundquist-number
regime. Fast reconnection is facilitated by the plasmoid instability.
The resultant reconnection rate is independent of
$S$ and is weakly dependent on the noise level. We have verified
the $S^{3/8}$ scaling of the number of plasmoids in the linear regime,
as predicted in Refs. \cite{LoureiroSC2007,BhattacharjeeHYR2009}.
In the nonlinear regime, the number of plasmoids follows a steeper
scaling and is proportional to $S$. We also have done statistical
studies of the local current sheets, and found that the current sheet
thickness and length both scale as $S^{-1}$, while the current density
scales as $S$. These findings are consistent with our heuristic argument
and the claim that the reconnection rate is independent of $S$ in
the high-$S$ regime. 

The fast reconnection rate we have obtained is approximately $0.01V_{A}B$,
which is similar to the values from other recent
resistive MHD studies,\cite{BhattacharjeeHYR2009,CassakSD2009}
but is smaller than the typical reconnection rate from collisionless
two-fluid or particle-in-cell simulations by an order of magnitude.
Which rate will be realized depends on how collisional the system
is. If the Sweet-Parker thickness $\delta_{SP}$ is greater than the
ion skin depth $d_{i}$ (or the ion Larmor radius at the sound speed,
$\rho_{s}$, if there is a guide field) in a system, a Sweet-Parker
layer is likely to form first. On the other hand, if $\delta_{SP}<d_{i}$
(or $\rho_{s}$), the reconnection would likely proceed dominated
by collisionless effects.\cite{Aydemir1992,MaB1996a,DorelliB2003,Bhattacharjee2004,CassakSD2005,CassakDS2007}
An interesting regime that has not drawn much attention before is
when $\delta_{SP}>d_{i}$ (or $\rho_{s}$) but $S>S_{c}$. Then we
expect the plasmoid instability to set in and the primary Sweet-Parker
layer will break into segments. This brings the thickness further
down to $\delta\sim\delta_{SP}(S_{c}/S)^{1/2}$. If $\delta>d_{i}$
(or $\rho_{s}$) then the system is still dominated by collisional
effects and we may end up getting a reconnection rate of $0.01V_{A}B$.
However, if $\delta<d_{i}$ (or $\rho_{s}$) then the system will
be in the collisionless regime. It is possible the reconnection rate
will be further enhanced. Indeed, the recent particle-in-cell simulation
by Daughton \emph{et al}. suggests this possibility \cite{DaughtonRAKYB2009}
but more needs to be done to determine if this is a general trend.
\begin{acknowledgments}
The authors would like to thank Dr. Brian P. Sullivan, Prof. Kai Germaschewski,
Dr. William Fox, Dr. Hongang Yang, Prof. Barrett N. Rogers, and Prof.
Chung-Sang Ng for beneficial conversations. We also acknowledge an
anonymous referee for many constructive suggestions. This work is
supported by the Department of Energy, Grant No. DE-FG02-07ER46372,
under the auspice of the Center for Integrated Computation and Analysis
of Reconnection and Turbulence (CICART) and the National Science Foundation,
Grant No. PHY-0215581 (PFC: Center for Magnetic Self-Organization
in Laboratory and Astrophysical Plasmas). Computations were performed
on facilities at National Energy Research Scientific Computing Center
and the Zaphod Beowulf cluster, which was funded in part by the Major
Research Instrumentation program of the National Science Foundation,
Grant No. ATM-0424905.
\end{acknowledgments}

\bibliographystyle{apsrev}
\bibliography{/Users/yop/Mydoc/reference/fullref}

\end{document}